\documentstyle[preprint,eqsecnum,aps,epsf]{revtex}	% preprint format

\newif\iftightenlines\tightenlinesfalse
\tightenlines\tightenlinestrue

\begin{document}
%
%%%%%%%%%%%%%%%%%%%%%%%%%%%%%%%%%%%%%%%%%%%%%%%%%%%%%%%%%%%%%%%%%%%%%%%%%%%%%%%
\def\mol{Mol}
\def\eslt{E\llap/_T}
\def\esl{E\llap/}
\def\msl{m\llap/}
\def\to{\rightarrow}
\def\te{\tilde e}
\def\tmu{\tilde\mu}
\def\ttau{\tilde\tau}
\def\tl{\tilde\ell}
\def\ttau{\tilde \tau}
\def\tg{\tilde g}
\def\tnu{\tilde\nu}
\def\tell{\tilde\ell}
\def\tq{\tilde q}
\def\tb{\tilde b}
\def\tst{\tilde t}
\def\tt{\tilde t}
\def\tw{\widetilde W}
\def\tz{\widetilde Z}

\hyphenation{mssm}
%\def\ds{\displaystyle}
%\def\ts{${\strut\atop\strut}$}
%
%%%%%%%%%%%%%%%%%%%% TITLE PAGE %%%%%%%%%%%%%%%%%%%%%%%%%%%%%%%%%%%%%%%%%%%%%%
\preprint{\vbox{\baselineskip=14pt%
   \rightline{FSU-HEP-950818}\break
}}
\title{COSMOLOGICAL RELIC DENSITY FROM \\
MINIMAL SUPERGRAVITY WITH IMPLICATIONS \\
FOR COLLIDER PHYSICS}
\author{Howard Baer and Michal Brhlik}
\address{
Department of Physics,
Florida State University,
Tallahassee, FL 32306 USA
}
\date{\today}
\maketitle
\begin{abstract}

Working within the framework of the minimal supergravity model with
gauge coupling unification and radiative electroweak symmetry breaking,
we evaluate the cosmological relic density from lightest neutralinos
produced in the early universe. Our numerical calculation is distinct in that
it involves direct evaluation of neutralino annihilation cross sections
using helicity amplitude techniques, and thus avoids the usual
expansion as a power series in terms of neutralino velocity.
Thus, our calculation includes relativistic Boltzmann averaging, neutralino
annihilation threshold effects, and proper treatment of integration
over Breit-Wigner poles. We map out regions
of parameter space that give rise to interesting cosmological dark matter
relic densities. We compare these regions with recent calculations of
the reach for supersymmetry by LEP2 and Tevatron Main Injector era
experiments. The cosmologically favored regions overlap considerably with the
regions where large trilepton signals are expected to occur at the Tevatron.
The CERN LHC $pp$ collider can make a thorough exploration
of the cosmologically favored region via gluino and squark searches.
In addition, over most of the favored region,
sleptons ought to be light enough to be detectable at both LHC and at a
$\sqrt{s}=500$ GeV $e^+e^-$ collider.
\end{abstract}

\medskip
\pacs{PACS numbers: 14.80.Ly, 98.80.Cq, 98.80.Dr}
%{\tt$\backslash$\string pacs\{\}}

%\narrowtext

%%%%%%%%%%%%%%%%%% MAIN TEXT %%%%%%%%%%%%%%%%%%%%%%%%%%%%%%%%%%%%%%%%%%%%%%%

\section{Introduction}

The Minimal Supersymmetric Standard Model\cite{MSSM} (MSSM) is one of the
leading candidate theories for physics beyond the Standard Model (SM). The MSSM
is a globally supersymmetric version of the SM, where supersymmetry
breaking is implemented by the explicit introduction of soft-supersymmetry
breaking terms. The MSSM is minimal in the sense that the fewest number of
additional new particles and interactions are incorporated which are
consistent with phenomenology.
In particular, possible baryon ($B$) and lepton ($L$)
number violating interactions are excluded from the superpotential
(the presence of {\it both} $B$ and $L$ violating interactions can lead to
catastrophic proton decay rates).
As a result, there exists a
conserved $R$-parity, where the multiplicative quantum number $R=+1$ for
ordinary particles, and $R=-1$ for superpartners.
A consequence of $R$-parity
conservation is that the lightest supersymmetric particle (LSP) is
absolutely stable.
Theoretical prejudice coupled with experimental constraints
strongly favor a color and charge neutral LSP\cite{WOLF}.
In addition, in the MSSM, the
LSP is strongly favored to be the massive, weakly interacting
lightest neutralino $\tz_1$\cite{BDT,JELLIS}.
$\tz_1$'s, if they exist,  would have been abundantly produced
in the early universe; if so, then relic neutralinos could well
make up the bulk of the dark matter in the universe today\cite{KT,JKG}.

In this paper, we restrict ourselves to the framework of the low energy
effective Lagrangian which is expected to result from, for instance,
supergravity grand unified models\cite{ARN}.
In these models, it is assumed that supersymmetry is broken in a hidden sector
of the theory. Supersymmetry breaking is then communicated to the
observable sector via gravitational interactions, leading to a common
mass $m_0$ for all scalar particles, a common mass $m_{1/2}$ for all
gauginos, a common trilinear coupling $A_0$, and a bilinear coupling $B_0$.
These soft supersymmetry breaking parameters are induced at energy scales
at or beyond the unification scales, but with (theoretically motivated)
values typically in the range $100-1000$ GeV.
The resulting theory, the MSSM with universal soft breaking terms,
is then regarded as an effective
theory with Lagrangian parameters renormalized at
an ultra-high scale $M_X \sim M_{GUT}-M_{Planck}$, and valid only
below this scale.
The corresponding weak scale sparticle couplings and masses can
then be calculated by evolving 26 parameters via
renormalization group equations\cite{RGE}
from the unification scale to the weak scale. An elegant
by-product\cite{RAD} of this mechanism
is that one of the Higgs boson mass squared terms is driven negative,
resulting in a breakdown of electroweak symmetry.
The radiative electroweak symmetry breaking constraint allows one to
essentially eliminate $B$ in favor of $\tan\beta$ (the ratio of Higgs field
vev's), and to calculate the magnitude of the superpotential Higgs mixing
term $\mu$ in terms of $M_Z$ (where we actually minimize the full one-loop
effective potential).
The model is then specified by only four
SUSY parameters (in addition to SM masses and couplings).
A hybrid set consisting of the common GUT scale scalar mass $m_0$,
common gaugino mass $m_{1/2}$, common SUSY-breaking trilinear
coupling $A_0$, along with the weak scale value of $\tan\beta$
proves to be a convenient choice. In addition, the sign of $\mu$ must
be stipulated.
These parameters fix the weak scale masses and couplings of all the
sparticles\cite{SPECTRA}.

The matter density of the universe $\rho$ is usually
parametrized\cite{KT,JKG} in terms of $\Omega=\rho /\rho_c$,
where $\rho_c ={3H_0^2/8\pi G_N} \simeq 1.88\times 10^{-29}h^2$
g/cm$^3$, and $h$, the Hubble scaling constant, is related to the Hubble
constant $H_0$ by $H_0=100h$ km/sec/Mpc. Here $h$ parametrizes our ignorance
of the true value of $H_0$, so that $0.5\alt h\alt 0.8$. Measurements of
galactic
rotation curves suggest $\Omega\sim 0.03 - 0.1$, compared to a luminous matter
density of $\Omega_{lum.}\alt 0.01$. Galactic clustering and galactic flows
suggest even larger values of $\Omega\sim 0.2-1$. Finally, the
theoretically attractive inflationary cosmological models require a flat
universe with $\Omega=1$. Meanwhile, estimates of the baryonic contribution
to the matter density of the universe from Big-Bang nucleosynthesis
suggest that $\Omega_{baryonic}\sim 0.01-0.1$. These analyses and estimates
suggest that the bulk of matter in the universe is (non-baryonic) dark matter.
Finally, analyses of structure formation in the universe in light of the COBE
measurements of anisotropies in the cosmic microwave background radiation
suggest that dark matter may be made of $\sim 60\%$ cold-dark matter
(weakly interacting massive particles or WIMPS, such as the lightest neutralino
$\tz_1$), $\sim 30\%$ hot dark matter (such as relic neutrinos), and
$\sim 10\%$ baryonic matter. This is the so-called
``mixed dark matter '' scenario.

The central idea\cite{KT} behind relic density calculations is that in the very
early
universe, neutralinos were being created and annihilated, but that they
were in a state of thermal equilibrium with the cosmic soup. As the universe
expanded and cooled, temperatures dropped low enough that neutralinos could
no longer be produced ($T\alt m_{\tz_1}$),
although they could still annihilate with one
another, at a rate governed by the total neutralino pair annihilation
cross section, and the neutralino number density.
Ultimately, as the universe expanded further, the expansion rate
outstripped the annihilation rate, thus freezing out the remaining neutralino
population of the universe, and locking in a neutralino relic density.
Our goal in this paper is to carry out estimates of the neutralino relic
density expected from the minimal SUGRA model. One solid constraint on
supersymmetric models with relic dark matter particles comes from the age
of the universe, which ought to be greater than $10$ ($15$) Gyrs; this
implies $\Omega h^2< 1$ ($0.25$). Thus, models with too large a relic
density would yield too young of a universe, in violation at least
with the age of the oldest stars in globular clusters. Furthermore,
models with $\Omega h^2< 0.025$ would not even be able to account for the dark
matter needed to explain galactic rotation: such models would be considered
cosmologically uninteresting. Models with intermediate values of
$0.025\alt \Omega h^2\alt 1$ are considered cosmologically interesting,
as they might explain galactic rotation and clustering, or might even
make up the matter density needed for inflationary cosmology, given a cold-dark
matter (CDM: $\Omega h^2\sim 0.25-0.64$) or mixed hot/cold dark matter
scenario (MDM: $\Omega h^2\sim .15-.4$).

Following the procedures outlined by Lee and Weinberg\cite{LW}, many groups
have calculated the relic neutralino
abundance\cite{HAIM,ELLIS,GRIEST,OLIVE,DREES,LOPEZ,ROSZ,AN}.
Early works involved calculating the most important
neutralino annihilation channels, usually assuming the LSP was a
photino. Later studies included various improvements, including
more annihilation channels, more general neutralino mixings, and more
realistic supersymmetric particle spectra.
A common thread amongst many papers was the calculation of
the Boltzmann-averaged
quantity $\sigma\times v$ using a power series expansion in
velocity. Such an approach was shown to be inaccurate when relativistic
effects were important, when annihilation proceeded through $s$-channel
poles, when threshold effects were important, or when
co-annihilation processes occured\cite{GS,GG}.
Many recent calculations have included some or all of these effects.

We have several goals in mind for the present paper.
\begin{itemize}
\item We wish to present reliable calculations for the neutralino relic
density in supersymmetric models. To this end we evaluate {\it all} $2\to 2$
neutralino annihilation diagram amplitudes numerically as complex numbers,
without approximation.
We perform Boltzmann averaging using the Gondolo-Gelmini formalism\cite{GG}.
This takes into account relativistic thermal averaging, while our numerical
helicity amplitude technique avoids the usual uncertainties inherent in
the velocity expansion, so that Breit-Wigner poles and threshold effects
are fully accounted for. Co-annihilation can occur when the two lightest
superpartners are very close in mass-- this situation rarely occurs within the
SUGRA framework adopted in this paper, and hence we ignore it.
\item We present results in the well-motivated SUGRA framework, which
includes gauge coupling unification, Higgs mass radiative
corrections\cite{HIGGS},
and radiative electroweak symmetry breaking using the one-loop effective
potential\cite{GRZ}.
\item Our results for the relic density calculation can be directly
related to recent calculations for various supersymmetry signals
expected at the LEP2\cite{LOPLEP,LEP2}, Tevatron\cite{BKT,LOPTEV,BCKT,MRENNA}
and LHC colliders\cite{BCPT,CHEN}.
In particular,
relic density calculations have a preference for light sleptons.
Such light sleptons may well be observable at LEP2 or LHC colliders, and
yield enhanced rates for $\tw_1\tz_2\to 3\ell$ states at the Tevatron
collider\cite{BT}.
\end{itemize}

To accomodate these goals, we present in Sec. II various details of our
relic density calculation, including those peculiar to the present approach.
In Sec. III, we present numerical results for our relic density calculations
in the $m_0\ vs.\ m_{1/2}$ plane of the minimal SUGRA model. In Sec. IV,
these results are explicitly compared to expectations for minimal
SUGRA at various
collider experiments, as worked out in a series of previous papers.
Finally, in Sec. V we present an overview and some conclusions.

\section{Calculational details}

We begin our determination of the neutralino relic density from minimal
supergravity by selecting a point in the SUGRA parameter space
\begin{eqnarray}
m_0,\ m_{1/2},\ \tan\beta ,\ A_0\ {\rm and}\ sign(\mu ),
\end{eqnarray}
where in addition we take the top quark mass $m_t=170$ GeV. The 26
renormalization group equations are iteratively run between the weak scale
and the GUT scale, which is defined as the point where the $U(1)$, $SU(2)$
and $SU(3)$ gauge couplings unify, and is typically $M_X\sim 2\times 10^{16}$
GeV. We use 2-loop RGE equations for gauge and Yukawa couplings
(with SUSY particle threshold effects), but only
1-loop equations for the running of the various soft breaking terms. The
1-loop effective Higgs potential is minimized to enforce radiative electroweak
symmetry breaking. Our procedure has been described in more detail in
Ref. \cite{BCMPT}, and has been incorporated into the event generator
ISAJET\cite{ISAJET}. At this point, a correlated sparticle mass spectrum
and couplings emerge from our input point in SUGRA parameter space.

The next step in our computation, after obtaining the superparticle spectrum,
is to evaluate the neutralino relic density by solving the Boltzmann
equation as formulated for a Friedmann-Robertson-Walker
cosmology\cite{KT}. Central to the evaluation of the relic density is the
computation of the fully relativistic,
thermally averaged neutralino annihilation cross section
times velocity, defined as
\begin{eqnarray}
<\sigma v_{\mol}>(T)={
{\int \sigma v_{\mol}e^{-E_1/T} e^{-E_2/T} d^3p_1 d^3p_2}\over
{\int e^{-E_1/T} e^{-E_2/T} d^3p_1 d^3p_2}
},
\end{eqnarray}
where $p_1$ ($E_1$) and $p_2$ ($E_2$) are the momentum and energy of the two
colliding particles in the cosmic, co-moving frame of reference, and $T$
is the temperature.
The above expression has been reduced to a one-dimensional integral by
Gondolo and Gelmini\cite{GG}, which yields
\begin{eqnarray}
<\sigma v_{\mol}>(x)={1\over{4xK_2^2({1\over x})}}
\int_2^{\infty} da \sigma (a) a^2(a^2-4) K_1({a\over x}),
\end{eqnarray}
where $x={T\over m_{\tz_1}}$, $a={\sqrt{s}\over m_{\tz_1}}$, $\sqrt{s}$ is
the subprocess energy, and $K_i$ are modified Bessel functions of order $i$.

We evaluate the neutralino annihilation cross section for
$\tz_1\tz_1\to f_1 f_2$ as
\begin{eqnarray}
d\sigma (a)={1\over {32\pi s}}
{{\lambda^{1\over 2}(s,m_{f_1}^2,m_{f_2}^2)}\over
{\lambda^{1\over 2}(s,m_{\tz_1}^2,m_{\tz_1}^2)}}
{\overline\Sigma} |{\cal M}|^2 d\cos\theta ,
\end{eqnarray}
where ${\overline\Sigma}|{\cal M}|^2$ is the spin summed and averaged
squared matrix element. Our calculation of the relic density is distinct in
that we evaluate ${\cal M}$ for {\it all} Feynman diagrams listed in Table 1
as complex numbers, using the HELAS\cite{HELAS} helicity amplitude
subroutine package. Thus, our approach avoids the usual uncertainties
associated with the expansion of cross section in terms of a power series in
velocity. The integration over $\cos\theta$ is performed numerically using
Gaussian quadratures.

To evaluate the neutralino relic density, the freeze out temperature $x_F$
is needed. The standard procedure here to iteratively solve the
freeze out relation
\begin{eqnarray}
x_F^{-1}=\log \Big[ {m_{\tz_1}\over {2\pi^3}} {\sqrt{45\over {2g_* G_N}}}
<\sigma v_{\mol}>_{x_F} x_F^{1\over 2}\Big],
\end{eqnarray}
by starting with a trial value $x_F={1\over 20}$. In the above, $g_*$ is the
effective number of degrees of freedom at $T=T_F$ ($\sqrt{g_*}\simeq 9$),
and $G_N$ is Newton's constant.

Finally, the relic density can be calculated from
\begin{eqnarray}
\Omega h^2= {\rho (T_0)\over {8.0992\times 10^{-47}\ {\rm GeV}^4}},
\end{eqnarray}
where
\begin{eqnarray}
\rho (T_0)\simeq 1.66\times{1\over M_{Pl}}
({{T_{m_{\tz_1}}}\over{T_\gamma}})^3
T_\gamma^3 {\sqrt{g_*}} {1\over {\int_0^{x_F}<\sigma v_{\mol}> dx}}.
\end{eqnarray}
To evaluate the integral in the above expression, we expand the modified
Bessel functions in Eq. 2.3 as power series in $x$, and then integrate over
$x$.
The result is
\begin{eqnarray}
\int_0^{x_F}<\sigma v_{\mol}> dx ={1\over {8\pi}} \int_2^\infty da \sigma (a)
a^{3\over 2} (a^2-4) F(a),
\end{eqnarray}
where
\begin{eqnarray*}
F(a)&=&\sqrt{{\pi\over {a-2}}} \left\{ 1-Erf(\sqrt{{{a-2}\over
x_F}})\right\}+\\
& &2({3\over 8a}-{15\over 4})\left\{ \sqrt{x_F} e^{-{{a-2}\over x_F}}-
\sqrt{\pi (a-2)} (1-Erf(\sqrt{{{a-2}\over x_F}}))\right\}+ \\
& & {2\over 3}({285\over 32}-{45\over 32a}-{15\over 18a^2})\times\\
& &\left\{ e^{-{{a-2}\over x_F}}\left[ x_F^{3\over 2}-2(a-2)\sqrt{x_F}\right]
+2\sqrt{\pi }(a-2)^{3\over 2}(1-Erf(\sqrt{{a-2}\over x_F} ))\right\} .
\end{eqnarray*}
In the above, virtually all the contribution to the integral comes from
$x<2.5$. We integrate the above expression numerically with Gaussian
quadratures, taking care to scan finely the regions with a Breit-Wigner pole.
In the region of a pole, the domain of integration must be broken into very
tiny
intervals, and obtaining convergence for a single point in parameter space
can take up to several hours of CPU time on a DEC ALPHA.

\section{Results from relic density calculation}

Our first numerical results for the relic density from minimal SUGRA models
are given in Fig. 1, where we plot contours of the neutralino relic
density $\Omega h^2$ in the $m_0\ vs.\ m_{1/2}$ parameter plane, where
we take $A_0=0$, $\tan\beta =2$, $\mu <0$ and $m_t=170$ GeV.
Changes in the $A_0$ parameter mainly affect 3rd generation sparticle
masses, and consequently result in only small changes in the relic density.
The regions labelled TH are excluded by theoretical considerations:
either there is a charged or colored LSP (or the $\tnu$ is LSP), or
the radiative electroweak symmetry breaking constraint breaks down.
The region labelled by EX corresponds to parameter space already excluded
by SUSY searches at LEP or Fermilab Tevatron experiments\cite{BCMPT}.

In almost all of
the plane, we find $\Omega h^2 >0.025$, {\it i.e.} large enough to explain the
galactic rotation curves. However, the region to the right of the
$\Omega h^2 =1$ contour is certainly excluded in that the age of the
universe would be younger than $10$ Gyrs. Meanwhile, a dominantly CDM
inflationary universe would lie in between the $\Omega h^2 =0.25-0.75$
contours. The COBE favored MDM inflationary universe would lie
between the $\Omega h^2 =0.15-0.4$ contours. For this latter favored region,
$m_{1/2}$ is bounded by $m_{1/2}\alt 400$ GeV (corresponding to
$m_{\tg}\alt 1000$ GeV), and $m_0<150$ GeV, unless the gluino is very
light ($m_{\tg}\simeq 300$ GeV).
(For comparison, various SUSY particle mass contours
for the same parameter choices are listed in Refs. \cite{BCMPT,BCKT,BCPT}.)
We find in general that large values of $m_0\agt 350$ GeV
(corresponding to $m_{\tell}\agt 250$ GeV) yield too young
a universe (due to suppression of $t$-channel slepton exchange diagrams),
except for the two narrow corridors in the lower right region
of the figure. The upper of the two corridors corresponds to neutralino
annihilation through the $Z$ pole, so the relic density is largely reduced
by $Z$ mediated $s$-channel annihilation diagrams. The lower of the two
corridors corresponds to annihilation through an $s$-channel light Higgs
$h$ pole-- in this case, the relic density falls rapidly to values even
below $\Omega h^2\sim 0.025$.

A qualitative feel for the relative importance of different annihilation
channels can be gleaned from Fig. 2. Here we plot for $m_0$ fixed at
200 GeV, as a function of $m_{1/2}$, the thermally averaged
annihilation cross section times velocity, integrated over temperature,
which enters into the relic density calculation (Eq. 2.8).
Larger cross sections correspond to smaller relic densities. As $m_{1/2}$
increases, the first pole we come to is annihilation via $s$-channel $h$,
where $\tz_1\tz_1\to b\bar b$ dominates. In these plots, $m_{\tz_1}$ scales
with $m_{1/2}$, and at the Higgs pole in this plot (on the edge of exclusion
by LEP Higgs search experiments),
$m_{\tz_1}\simeq 30$ GeV and $m_{\tw_1}\simeq 70$ GeV.
As one moves to higher $m_{1/2}$,
annihilation through the $Z$ pole is reached, which is dominated by
$\tz_1\tz_1\to d\bar d,\ s\bar s$, and $b\bar b$.
For values of $m_{1/2}$ away from poles,
annihilation via $t$-channel slepton and sneutrino exchange dominates.
For even higher
$m_{1/2}$ values, annihilation into channels such as $hh,\ Zh,\ WW$ and $ZZ$
open up, but never dominate for the parameter choices in this plot.
Annihilation into other channels such as $HA$, $AA$, $HH$ amd $H^+H^-$
are included in our calculation, but unimportant given our SUGRA sparticle
mass spectrum, which yields very large masses for Higgs bosons other than $h$.
The onset of the $\tz_1\tz_1\to t\bar t$ can be detected in the
$\tz_1\tz_1\to\Sigma u_i\bar{u_i}$ curve around $m_{1/2}\sim 400$ GeV.

If we plot the relic density for the same parameter choices, but flip the
sign of $\mu$, so that $\mu >0$, then we obtain the results of Fig. 3. The
relic density contours in this case are similar to those of Fig. 1 for
large values of $m_{1/2}$, where annihilation dominantly occurs via
slepton exchange. The kink in the contours is due to the onset of the
$\tz_1\tz_1\to t\bar t$ channel.
In this case, annihilation through $t$-channel $\tst_1$ exchange makes
a large contribution to the total annihilation cross section.
For smaller values of $m_{1/2}$,
in contrast to Fig. 1, we find only one
corridor extending to large $m_0$ where the relic density drops to
cosmologically un-interesting values. In this case, the $Z$ and $h$ poles
very nearly overlap for $m_{\tz_1}\sim 46$ GeV. This can be seen in more detail
in Fig. 4, where again we show the thermally averaged cross section versus
$m_{1/2}$, for $m_0 =200$ GeV.

Finally, we show again the neutralino relic density $\Omega h^2$ in the
$m_0\ vs.\ m_{1/2}$ plane for the same parameter choices as Fig. 1, except
now we take a large value of $\tan\beta =10$. For this case, we note the
rather broad band at $m_{1/2}\sim 100-140$ GeV, where $\Omega h^2 <0.025$-
too low to explain even the galactic rotation curves, and due again to
annihilation through the $s$-channel graphs. In fact, inflationary models,
which require $\Omega h^2 \agt 0.15$, are only allowed if $m_{1/2}>150$ GeV,
corresponding to $m_{\tg}>400$ GeV. In this plot, there is a significant
region extending to large values of $m_0$, corresponding to large $m_{\tq}$
and large $m_{\tell}$, for $m_{1/2}\sim 150-190$ GeV. The contributing
thermally averaged subprocess cross sections are again shown in Fig.~6 for
$m_0=200$ GeV. In this plot, the $Z$ pole annihilation channel occurs at
$m_{1/2}\simeq 110$ GeV, followed by the Higgs pole at $m_{1/2}\simeq 130$ GeV.
The rough overlap of these two pole contributions leads to the single broad
corridor of low $\Omega h^2$ shown in Fig.  5.

\section{Implications for SUSY searches at colliders}

Recently, various papers have been written on the prospects for
supersymmetry at the LEP2 $e^+e^-$ collider\cite{LOPLEP,LEP2}, the Fermilab
Tevatron $p\bar p$ collider\cite{BT,BKT,LOPTEV,BCKT,MRENNA} and
the CERN LHC $pp$ collider\cite{BCPT,CHEN}.
Our objective in this section is to assess the prospects
for discovery of SUGRA at hadron and $e^+e^-$ colliders, given the additional
constraints from requiring a reasonable value for the neutralino relic
density. We mainly focus on
the collider results of Refs. \cite{LEP2,BKT,BCKT,BCPT,CHEN,DPF},
since they were performed
in a consistent framework, in the same $m_0\ vs.\ m_{1/2}$ plane.

In Fig. 7, we again show the neutralino relic density contours in the
$m_0\ vs.\ m_{1/2}$ parameter plane, for the same parameter choices as in
Fig. 1. In addition, we have added on contours for SUSY discovery at various
colliders. Supersymmetric particles ought to be discoverable at LEP2
operating at $\sqrt{s}=190$ GeV, with integrated luminosity
$\int {\cal L}dt=300$ fb$^{-1}$ below the contour labelled LEP2\cite{LEP2}.
The
lower-left bulge in the LEP2 contour is where sleptons ought to be detectable,
while beneath the contour running along $m_{1/2}\simeq 100$ GeV (which
runs through the neutralino $Z$-pole annihilation region), charginos
ought to be detectable. By comparing, we see that the region accessible by LEP2
generally has $\Omega h^2<0.15$ {\it i.e.} not the most cosmologically
favored region, but with enough dark matter to explain galactic rotation.
However, the contour labelled with LEP2-Higgs shows the reach of LEP2 for
the light SUSY Higgs boson $h$, which is just below $m_{1/2}\sim 400$ GeV.
This region completely encloses the favored MDM region. The implication is
that if MDM explains dark matter in the universe, and if $\tan\beta$ is small
and $\mu <0$, then LEP2 ought to discover at least the light SUSY Higgs boson.

The dashed contour labelled Tevatron is a composite of the reach of
Tevatron Main Injector era ($\sqrt{s}=2$ TeV; $\int {\cal L}dt=1$ fb$^{-1}$
integrated luminosity) experiments for multi-jet$+\eslt$ events\cite{BKT},
and mainly, for $\tw_1\tz_2\to 3\ell$ events\cite{BCKT}. We see that the
largest reach by Tevatron experiments occurs exactly in the
cosmologically favored MDM region, and can reach to $m_{1/2}\sim 160$ GeV,
corresponding to $m_{\tg}\sim 440$ GeV. This is no accident: a reasonable
neutralino annihilation cross section generally requires $m_{\tell}\alt 200$
GeV; these lighter sleptons give rise to enhanced leptonic decay of
neutralinos,
leading to large rates for $\tw_1\tz_2\to 3\ell$ events.
Since lower values of $m_{1/2}$
are preferred by fine-tuning arguments\cite{FT}, there is a good chance
Tevatron experiments could discover SUSY via $3\ell$ events if nature chose
this parameter set.

We also compare the results of Fig. 7 with expectations for supersymmetry
at the CERN LHC collider. Of course, LHC experiments can cover the whole
parameter plane up to $m_{1/2}\sim 600-800$ GeV with only $\int {\cal L}dt=10$
fb$^{-1}$ of integrated luminosity, at $\sqrt{s}=14$ TeV, via searches for
multi-jet$+\eslt$ events from gluino and squark cascade decays\cite{BCPT},
so discovery of SUSY would be no problem.
We also plot in Fig. 7 the contour beneath which
sleptons ought to be visible at LHC\cite{BCPT,CHEN}. We see that the
cosmologically favored MDM region falls almost entirely within the slepton
discovery region, so that if the MDM scenario is correct, then LHC has a very
high
probability to discover a slepton.
Since sleptons are relatively light, LHC experiments ought as well to be
sensitive to $\tw_1\tz_2\to 3\ell$ events over much, but not all, of the
favored MDM region\cite{BCPT,CHEN}.
(In some of the favored region, $\tz_2\to \nu\tnu$ or
$\tz_1 h$, thus spoiling the signal.)
Finally, since $m_{\tell}\alt 250$ GeV in the MDM scenario, sleptons would
then likely be visible at a linear $e^+e^-$ collider operating
at $\sqrt{s}=500$ GeV.

In Fig. 8, we show the same relic density contours as in Fig. 3 ($\tan\beta =2$
, $\mu >0$), and compare again with expectations for colliders. In this case,
we see the LEP2 contour again lies in a region of $\Omega h^2 <0.15$,
although it does encompass the cosmologically interesting region around
$(m_0,m_{1/2})\sim (100,110)$. The LEP2 Higgs contour in this case lies at
$m_{1/2}\sim 170$ GeV, and thus covers only a portion of the MDM favored
region.
Thus, if the MDM scenario is correct, and $\tan\beta$ is small, minimal SUGRA
sparticles or light Higgs boson might still not be accessible at LEP2.
We also plot the contour due to the combined Tevatron MI reach. In this case,
there is a large Tevatron reach due to $\tw_1\tz_2\to 3\ell$
extending to $m_{1/2}\sim 230$ GeV, overlapping considerably with
the MDM region. Finally, we note once again that LHC can cover the whole plane
via multi-jet$+\eslt$ searches. In addition, the MDM region lies again almost
entirely within the LHC slepton search region, and overlaps substantially
with the LHC $\tw_1\tz_2\to 3\ell$ clean trilepton region\cite{CHEN}.

Last of all, we turn to Fig. 9, which compares the neutralino relic
density with collider search regions for large $\tan\beta =10$, with $\mu <0$.
In this case, we note that the MDM favored region lies entirely above
the region that is searchable at LEP2. In addition, for this case, the lightest
Higgs boson has mass $m_h \agt 90$ GeV throughout the plane, beyond the
reach of LEP2 at $\sqrt{s}=190$ GeV. Hence, if $\tan\beta$ is large, and the
MDM scenario is correct, then there would be little hope of seeing SUSY at
LEP2.
In this case, the prospect for minimal SUGRA at Tevatron MI is even worse,
except for the narrow region extending along $m_0\sim 100$ GeV, which enters
into the cosmologically favored MDM region. Finally, we note that once again
the LHC slepton reach contour excloses most of the MDM region, with the main
exception being the band of allowed MDM region extending to large $m_0$
along $m_{1/2}\sim 160-170$ GeV. The LHC $\tw_1\tz_2\to 3\ell$ region
encloses pieces of the MDM region, but leaves significant
areas uncovered\cite{CHEN}.

\section{Conclusion}

In this paper, working within the minimal supergravity model with
radiative electroweak symmetry breaking and universal GUT scale
soft supersymmetry breaking terms, we have evaluated the cosmological
relic density from neutralinos produced in the early universe. Our technique
was to evaluate {\it all} lowest order neutralino annihilation Feynman
diagrams as complex helicity amplitudes. We then performed the necessary
integrations numerically, preserving relativistic covariance, and avoiding the
usual expansion as a power series in velocity. While this approach might be
regarded as a brute force numerical calculation, it does include
relativistic thermal averaging, annihilation threshold effects, and careful
integration over Breit-Wigner poles. We do not include co-annihilation
processes in our calculations, which are however unimportant within the SUGRA
framework, in which we work.

Our numerical results for the neutralino relic density were presented in
Figures 1-6. For the favored mixed dark matter scenario, for which
$0.15<\Omega h^2 <0.4$, we find that, unless annihilation occurs via
$s$-channel
$Z$ or $h$ exchange (in which case $m_{\tg}<300-400$ GeV\cite{BDKNT}),
$m_{\tg}\alt 1000$ GeV, and $m_{\tell}\alt 250$ GeV. The less conservative
constraint from the age of the universe ($\Omega h^2 <1$) yields larger bounds
on sparticle masses.

We also examined the implications of our relic density calculations for
collider searches for the sparticles of minimal SUGRA. These results have
been summarized in Figs. 7-9. Within the MDM range of $\Omega h^2$, we find
that
LEP2 has a high probability to detect a light Higgs boson if
$\tan\beta$ is small and $\mu <0$. For the opposite sign of $\mu$, $m_h$ can be
larger, and detection at LEP2 is less certain. Prospects for detection
of sleptons or charginos at LEP2 are less bright: generally, if $m_{\tell}<90$
GeV, $t$-channel neutralino annihilation is too large, leading to rather
low values of neutralino relic density. Likewise, if $m_{\tw_1}<90$ GeV, then
$m_{\tz_1}\alt 45$ GeV, and neutralinos can annihilate via $s$-channel $Z$
or $h$ exchange, again leading to only a small relic abundance.

Prospects for discovering SUGRA at Tevatron MI experiments are
somewhat brighter, since a reasonable relic density requires roughly
$100< m_{\tell}<250$ GeV. Such a slepton mass range generally leads to enhanced
leptonic decays of neutralinos, giving Tevatron experiments a good chance to
find SUGRA via $\tw_1\tz_2\to 3\ell$ searches.

The CERN LHC $pp$ collider can make a thorough search for supersymmetry
over all the allowed parameter space in the multi-jet $+\eslt$ channel.
However, the rather light slepton masses required for reasonable neutralino
relic densities falls within the range of LHC experimental sensitivity, so
there is a good chance to find sleptons at LHC if, for instance, the MDM
scenario turns out to be correct. Likewise, experiments at
an $e^+e^-$ linear collider
operating at $\sqrt{s}\sim 500$ GeV would stand a good chance of discovering
sleptons, since they would be sensitive to slepton pair production for
$m_{\tell}\alt 230$ GeV\cite{HIT,DPF}.

\smallskip
\noindent{\it Note added: Upon completion of this work, a preprint
appeared which addressed the neutralino dark matter relic density in
SUGRA models with non-universal soft-breaking terms\cite{NEWELLIS}.}

%%%%%%%%%%%%%%%%%%%%%%%%% ACKNOWLEDGEMENTS
%%%%%%%%%%%%%%%%%%%%%%%%%%%%%%%%%%%%%%%

%\newpage
\acknowledgments

We thank X. Tata and C. H. Chen for discussions, and X. Tata
for comments on the manuscript.
This research was supported in part by the U.~S. Department of Energy
under grant number DE-FG-05-87ER40319.

%%%%%%%%%%%%%%%%%%%%% REFERENCES %%%%%%%%%%%%%%%%%%%%%%%%%%%%%%%%%%%%%%%%%%%%%%
%

%
\newpage
%
%%%%%%%%%%%%%%%%%%%%%%%%%% TABLES %%%%%%%%%%%%%%%%%%%%%%%%%%%%%%%%%%%%%%%%%%%
%
\begin{table}
\caption[]{A tabulation of Feynman diagrams contributing to our
neutralino annihilation cross section calculation.}
\end{table}
\bigskip

\begin{tabular}{|p{1.5in}|p{1.5in}|p{1.5in}|p{1.5in}|} \hline
& \multicolumn{3}{c|}{ Particles exchanged } \\
\cline{2-4}
\multicolumn{1}{|c|}{Process} &
\hspace{.4in}{s-channel} &
\hspace{.4in} {t-channel} &
\hspace{.4in} {u-channel} \\ \hline
\hline
\hspace{.15in} $\tz_1\tz_1\rightarrow Z^{0}Z^{0}$ &
\hspace{.5in} $h,H$ &
\hspace{.5in} $\tz_{1,2,3,4}$ &
\hspace{.5in} $\tz_{1,2,3,4}$ \\ \hline
\hspace{.15in} $\tz_1\tz_1\rightarrow W^{+}W^{-}$ &
\hspace{.5in} $h,H,Z^{0}$ &
\hspace{.5in} $\tw_{1,2}^{\pm}$ &
\hspace{.5in} $\tw_{1,2}^{\pm}$ \\ \hline
\hspace{.15in} $\tz_1\tz_1\rightarrow Z^{0}h$ &
\hspace{.5in} $Z^{0},A$ &
\hspace{.5in} $\tz_{1,2,3,4}$ &
\hspace{.5in} $\tz_{1,2,3,4}$ \\ \hline
\hspace{.15in} $\tz_1\tz_1\rightarrow Z^{0}H$ &
\hspace{.5in} $Z^{0},A$ &
\hspace{.5in} $\tz_{1,2,3,4}$ &
\hspace{.5in} $\tz_{1,2,3,4}$ \\ \hline
\hspace{.15in} $\tz_1\tz_1\rightarrow Z^{0}A$ &
\hspace{.5in} $h,H$ &
\hspace{.5in} $\tz_{1,2,3,4}$ &
\hspace{.5in} $\tz_{1,2,3,4}$ \\ \hline
\hspace{.15in} $\tz_1\tz_1\rightarrow W^{-}H^{+}$ &
\hspace{.5in} $h,H,A$ &
\hspace{.5in} $\tw_{1,2}^{\pm}$ &
\hspace{.5in} $\tw_{1,2}^{\pm}$ \\ \hline
\hspace{.15in} $\tz_1\tz_1\rightarrow W^{+}H^{-}$ &
\hspace{.5in} $h,H,A$ &
\hspace{.5in} $\tw_{1,2}^{\pm}$ &
\hspace{.5in} $\tw_{1,2}^{\pm}$ \\ \hline
\hspace{.15in} $\tz_1\tz_1\rightarrow hh$ &
\hspace{.5in} $h,H$ &
\hspace{.5in} $\tz_{1,2,3,4}$ &
\hspace{.5in} $\tz_{1,2,3,4}$ \\ \hline
\hspace{.15in} $\tz_1\tz_1\rightarrow HH$ &
\hspace{.5in} $h,H$ &
\hspace{.5in} $\tz_{1,2,3,4}$ &
\hspace{.5in} $\tz_{1,2,3,4}$ \\ \hline
\hspace{.15in} $\tz_1\tz_1\rightarrow hH$ &
\hspace{.5in} $h,H$ &
\hspace{.5in} $\tz_{1,2,3,4}$ &
\hspace{.5in} $\tz_{1,2,3,4}$ \\ \hline
\hspace{.15in} $\tz_1\tz_1\rightarrow AA$ &
\hspace{.5in} $h,H$ &
\hspace{.5in} $\tz_{1,2,3,4}$ &
\hspace{.5in} $\tz_{1,2,3,4}$ \\ \hline
\hspace{.15in} $\tz_1\tz_1\rightarrow hA$ &
\hspace{.5in} $Z^{0},A$ &
\hspace{.5in} $\tz_{1,2,3,4}$ &
\hspace{.5in} $\tz_{1,2,3,4}$ \\ \hline
\hspace{.15in} $\tz_1\tz_1\rightarrow HA$ &
\hspace{.5in} $h,H$ &
\hspace{.5in} $\tz_{1,2,3,4}$ &
\hspace{.5in} $\tz_{1,2,3,4}$ \\ \hline
\hspace{.15in} $\tz_1\tz_1\rightarrow H^{+}H^{-}$ &
\hspace{.5in} $h,H,Z^{0}$ &
\hspace{.5in} $\tw_{1,2}^{\pm}$ &
\hspace{.5in} $\tw_{1,2}^{\pm}$ \\ \hline
\hspace{.15in} $\tz_1\tz_1\rightarrow f\bar f$ &
\hspace{.5in} $Z^{0},h,H,A$ &
\hspace{.5in} $\tilde f_{1,2}^{\pm}$ &
\hspace{.5in} $\tilde f_{1,2}^{\pm}$ \\ \hline
\end{tabular}

%%%%%%%%%%%%%%%%%%%%%% FIGURE CAPTIONS %%%%%%%%%%%%%%%%%%%%%%%%%%%%%%%%%%%%%%
\begin{figure}
\caption[]{Plot of contours of constant $\Omega h^2$ in the $m_0\ vs.\ m_{1/2}$
plane, where $A_0=0$, $\tan\beta =2$, $\mu <0$ and $m_t=170$ GeV.
The regions labelled by TH (EX) are excluded by theoretical (experimental)
considerations.
}
\end{figure}

\begin{figure}
\caption[]{Relativistic thermally averaged cross section times velocity
as a function of $m_{1/2}$, for $m_0=200$ GeV, with other parameters as in
Fig. 1.
}
\end{figure}

\begin{figure}
\caption[]{Same as Fig. 1, except now $\mu >0$.
}
\end{figure}

\begin{figure}
\caption[]{Relativistic thermally averaged cross section times velocity
as a function of $m_{1/2}$, for $m_0=200$ GeV, with other parameters as in
Fig. 3.
}
\end{figure}

\begin{figure}
\caption[]{Same as Fig. 1, except now $\tan\beta =10$.
}
\end{figure}

\begin{figure}
\caption[]{Relativistic thermally averaged cross section times velocity
as a function of $m_{1/2}$, for $m_0=200$ GeV, with other parameters as in
Fig. 5.
}
\end{figure}

\begin{figure}
\caption[]{Same as Fig. 1, except
we also show the region below the contour labelled MI,
which is accessible to Tevatron Main Injector era experiments, and the
region below the LEP2 contour, where sparticles are accessible
to experiments at LEP2 operating at $\sqrt{s}=190$ GeV. Below the LEP2-Higgs
contour, the lightest SUSY Higgs $h$ is accessible at LEP2. Below the LHC
contour, sleptons ought to be accessible to LHC experiments.
}
\end{figure}

\begin{figure}
\caption[]{Same as Fig. 7, except now $\mu >0$.
}
\end{figure}

\begin{figure}
\caption[]{Same as Fig. 7, except now $\tan\beta =10$.
}
\end{figure}


\begin{references}
%1
\bibitem{MSSM} For reviews of the MSSM, see
H.~P.~Nilles, Phys.~Rep. {\bf 110}, 1 (1984);
H. Haber and G. Kane, Phys.~Rep. {\bf 117}, 75 (1985);
X.~Tata, in {\it The Standard Model and Beyond},
p.~304, edited by J.~E.~Kim, World Scientific (1991).
%2
\bibitem{WOLF} S. Wolfram, Phys. Lett. {\bf B82}, 65 (1979);
C. B. Dover, T. K. Gaisser and G. Steigman,
Phys. Rev. Lett. {\bf 42}, 1117 (1979);
P. F. Smith {\it et. al.}, Nucl. Phys. {\bf B206}, 333 (1982);
E. Norman {\it et. al.}, Phys. Rev. Lett. {\bf 58}, 1403 (1987).
%
\bibitem{BDT} H. Baer, M. Drees and X. Tata, Phys. Rev. {\bf D41}, 3414 (1990).
%3
\bibitem{JELLIS} J. Ellis, D. Nanopoulos, L. Roszkowski and D. Schramm,
Phys. Lett. {\bf B245}, 251 (1990); L. Krauss,
Phys. Rev. Lett. {\bf 64}, 999 (1990).
%
\bibitem{KT} E.~W. Kolb and M. S. Turner, {\it The Early Universe},
(Addison-Wesley, Redwood City, 1989).
%4
\bibitem{JKG} G. Jungman, M. Kamionkowski and K. Griest, SU-4240-605 (1995),
(submitted to Physics Reports); see also J. Ellis, CERN-TH.7083/93 (1993).
%5
\bibitem{ARN} A. H. Chamseddine, R. Arnowitt and P. Nath,
Phys. Rev. Lett. {\bf 49}, 970 (1982);
R.~Arnowitt and P.~Nath, {\it Lectures presented at the VII J.~A.~ Swieca
Summer School, Campos do Jordao, Brazil, 1993} CTP-TAMU-52/93;
{\it Properties of SUSY
Particles}, L. Cifarelli and V.~Khoze, Editors, World Scientific (1993);
for a recent discussion, see {\it e.g.} M. Drees and S. Martin, in
{\it Electroweak Symmetry Breaking and Beyond the Standard Model},
edited by T. Barklow, S. Dawson, H. Haber and J. Siegrist, (World Scientific,
to be published).
%7
\bibitem{RGE} K. Inoue, A.~Kakuto, H.~Komatsu and H.~Takeshita, Prog. Theor.
Phys. {\bf 68}, 927 (1982) and {\bf 71}, 413 (1984).
%
\bibitem{RAD} L.~Iba\~nez and G.~Ross, Phys. Lett. {\bf B110}, 215 (1982);
L.~Iba\~nez, Phys. Lett. {\bf B118}, 73 (1982); J.~Ellis, D.~Nanopoulos and
K.~Tamvakis, Phys. Lett. {\bf B121}, 123 (1983); L. Alvarez-Gaum\'e,
J.~Polchinski and M.~Wise, Nucl. Phys. {\bf B121}, 495 (1983).
%
\bibitem{SPECTRA} Some recent analyses of supergravity mass patterns include,
J. Ellis and F. Zwirner, Nucl. Phys. {\bf B338}, 317 (1990);
G. Ross and R.~G.~Roberts, Nucl. Phys. {\bf B377}, 571 (1992); R.~Arnowitt and
P.~Nath, Phys. Rev. Lett. {\bf 69}, 725 (1992); M.~Drees and M.~M.~Nojiri,
Nucl. Phys. {\bf B369}, 54 (1993); S.~Kelley {\it et. al.},
%, J.~Lopez, D.~Nanopoulos, H.~Pois and K.~Yuan,
Nucl. Phys. {\bf B398}, 3 (1993);
M. Olechowski and S. Pokorski, Nucl. Phys. {\bf B404}, 590 (1993);
V.~Barger, M.~Berger and P.~Ohmann, {\it ibid.} {\bf 49}, 4908 (1994);
G. Kane, C.~Kolda, L.~Roszkowski and J.~Wells, Phys. Rev. {\bf D49}, 6173
(1994); D.~J.~Casta\~no, E.~Piard and P.~Ramond, Phys. Rev. {\bf D49}, 4882
(1994); W.~de~Boer, R.~Ehret and D.~Kazakov, Karlsruhe preprint,
IEKP-KA/94-05 (1994); H. Baer, C. H. Chen, R. Munroe, F. Paige and X. Tata,
Phys. Rev. {\bf D51}, 1046 (1995).
%5
\bibitem{LW} B. Lee and S. Weinberg, Phys. Rev. Lett. {\bf 39}, 165 (1977).
%
\bibitem{HAIM} H. Goldberg, Phys. Rev. Lett. {\bf 50}, 1419 (1983).
%
\bibitem{ELLIS} J. Ellis, J. Hagelin, D. Nanopoulos and M. Srednicki,
Phys. Lett. {\bf B127}, 233 (1983);
J. Ellis, J. Hagelin, D. Nanopoulos, K. Olive and M. Srednicki,
Nucl. Phys. {\bf B238}, 453 (1984); J. Ellis, J. Hagelin and D. Nanopoulos,
Phys. Lett. {\bf B159}, 26 (1985).
%
\bibitem{GRIEST} K. Griest, Phys. Rev. {\bf D38}, 2357 (1988);
K. Griest, M. Kamionkowski and M. Turner, Phys. Rev. {\bf D41}, 3565 (1990).
%
\bibitem{OLIVE} K. Olive and M. Srednicki, Phys. Lett. {\bf B230}, 78 (1989)
and Nucl. Phys. {\bf B355}, 208 (1991); J. MacDonald, K. Olive and M.
Srednicki,
Phys. Lett. {\bf B283}, 80 (1992).
%
\bibitem{DREES} M. Nojiri, Phys. Lett. {\bf B261}, 76 (1991);
M. Drees and M. Nojiri, Phys. Rev. {\bf D47}, 376 (1993).
%
\bibitem{LOPEZ} J. Lopez, D. Nanopoulos and K. Yuan,
Phys. Lett. {\bf B267}, 219 (1991) and Nucl. Phys. {\bf B370}, 445 (1992);
S. Kelley, J. Lopez, D. Nanopoulos, H. Pois and K. Yuan,
Phys. Lett. {\bf B273}, 423 (1991).
%
\bibitem{ROSZ} J. Ellis and L. Roszkowski, Phys. Lett. {\bf B283}, 252 (1992);
L. Roszkowski and R. Roberts, Phys. Lett. {\bf B309}, 329 (1993);
G. Kane, C. Kolda, L. Roszkowski and J. Wells, Ref. \cite{SPECTRA}.
%
\bibitem{AN} R. Arnowitt and P. Nath, Phys. Rev. Lett. {\bf 70}, 3696 (1993)
and CTP-TAMU-53/94 (1994).
%
\bibitem{GS} K. Griest and D. Seckel, Phys. Rev. {\bf D43}, 3191 (1991).
%
\bibitem{GG} P. Gondolo and G. Gelmini, Nucl. Phys. {\bf B360}, 145 (1991).
%
\bibitem{HIGGS} H. Haber and R. Hempfling, Phys. Rev. Lett. {\bf 66}, 1815
(1991);
J. Ellis, G. Ridolfi and F. Zwirner, Phys. Lett. {\bf B257}, 83 (1991);
T. Okada, H. Yamaguchi and T. Tanagida, Prog. Theor. Phys. Lett. {\bf 85},1
(1991);
we use the calculations of M. Bisset, U. of Hawaii Ph. D. thesis,
UH-511-813-94 (1994), as implemented in ISAJET\cite{ISAJET}.
%
\bibitem{GRZ} G. Gamberini, G. Ridolfi and F. Zwirner,
Nucl. Phys. {\bf B331}, 331 (1990).
%
\bibitem{LOPLEP} J.~Lopez, D.~Nanopoulos, H.~Pois, X.~Wang and A.~Zichichi,
Phys. Rev. {\bf D48}, 4062 (1993).

\bibitem{LEP2} H. Baer, M. Brhlik, R. Munroe and X. Tata, FSU-HEP-950501
(1995).
%
\bibitem{BKT} H.~Baer, C.~Kao and X.~Tata, Phys. Rev. {\bf D48}, R2978 (1993)
and Phys. Rev. {\bf D51}, 2180 (1995).
%
\bibitem{LOPTEV} J. Lopez, D. Nanopoulos, X.~Wang and A.~Zichichi, Phys. Rev.
{\bf D48}, 2062 (1993) and CERN-TH.7535/94;
J. Lopez, D. Nanopoulos, G.~Park, X.~Wang and A.~Zichichi,
Phys. Rev. {\bf D50}, 2164 (1994); T. Kamon, J.~Lopez, P.~McIntyre and
J.~T.~White, Phys. Rev. {\bf D50}, 5676 (1994).
%
\bibitem{BCKT} H. Baer, C. H. Chen, C. Kao and X. Tata,
Phys. Rev. {\bf D52}, 1565 (1995).
%
\bibitem{MRENNA} S. Mrenna, G. Kane, G. Kribs and J. Wells, UM-TH-95-14 (1995).
%
\bibitem{BCPT} H. Baer, C. H. Chen, F. Paige and X. Tata,
Phys. Rev. {\bf D49}, 3283 (1994), Phys. Rev. {\bf D50}, 4508 (1994) and
FSU-HEP-950204 (1995), Phys. Rev. {\bf D} (in press).
%
\bibitem{CHEN} C. H. Chen, Florida State University Ph. D. thesis
FSU-HEP-950720 (1995).
%
\bibitem{BT} H. Baer and X. Tata, Phys. Rev. {\bf D47}, 2739 (1993).
%
\bibitem{BCMPT} See H. Baer, C. H. Chen, R. Munroe, F. Paige and X. Tata,
Ref. \cite{SPECTRA}.
%
\bibitem{ISAJET}  F. Paige and S. Protopopescu, in {\it Supercollider Physics},
p. 41, ed.\ D. Soper (World Scientific, 1986);
H. Baer, F. Paige, S. Protopopescu and X. Tata, in
{\it Proceedings of the Workshop on Physics at Current Accelerators
and Supercolliders}, ed.\ J. Hewett, A. White and D. Zeppenfeld,
(Argonne National Laboratory, 1993).
%
\bibitem{HELAS} HELAS: HELicity Amplitude Subroutines for Feynman Diagram
Evaluations, H. Murayama, I. Watanabe and K. Hagiwara, KEK-91-11 (1992).
%
\bibitem{FT} R.~Barbieri and G.~Giudice, Nucl. Phys. {\bf B306}, 63 (1988);
G.~Anderson and D.~Casta\~no, MIT-CTP-2369 (1994) presents an updated
measure of fine-tuning, and associated bounds on sparticle masses.
%
\bibitem{BDKNT} H.~Baer, M.~Drees, C.~Kao, M.~Nojiri and X.~Tata,
Phys. Rev. {\bf D50}, 2148 (1994).
%
\bibitem{HIT} T.~Tsukamoto, K.~Fujii, H.~Murayama, M.~Yamaguchi and Y. Okada,
Phys. Rev. {\bf D51}, 3153 (1995).
%
\bibitem{DPF} H. Baer {\it et. al.}, in {\it Electroweak Symmetry Breaking
and Beyond the Standard Model},
edited by T. Barklow, S. Dawson, H. Haber and J. Siegrist, (World Scientific,
to be published), FSU-HEP-950401.
%
\bibitem{NEWELLIS} V. Berezinsky {\it et. al.}, CERN-TH 95-206 (1995).
%
\end{references}
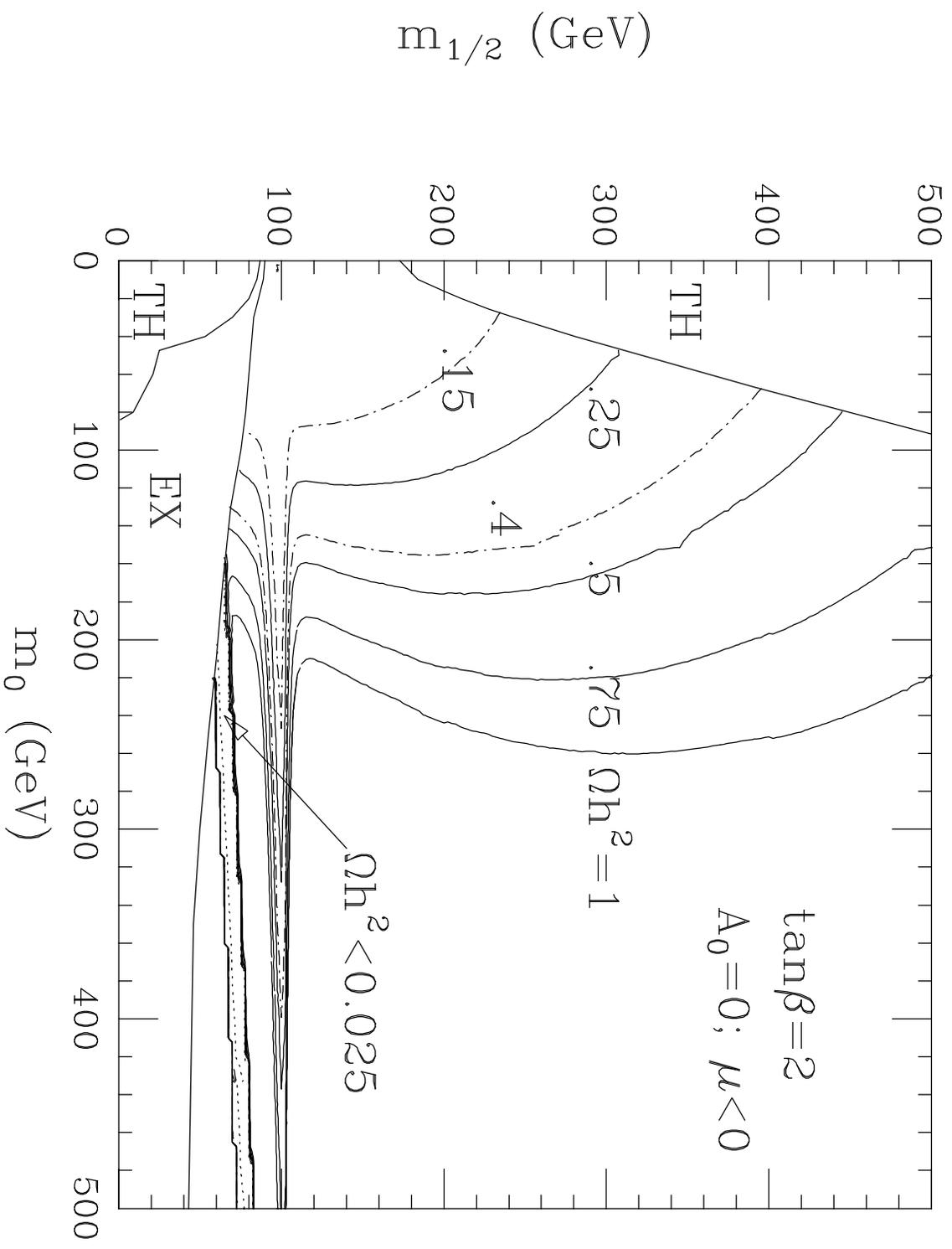
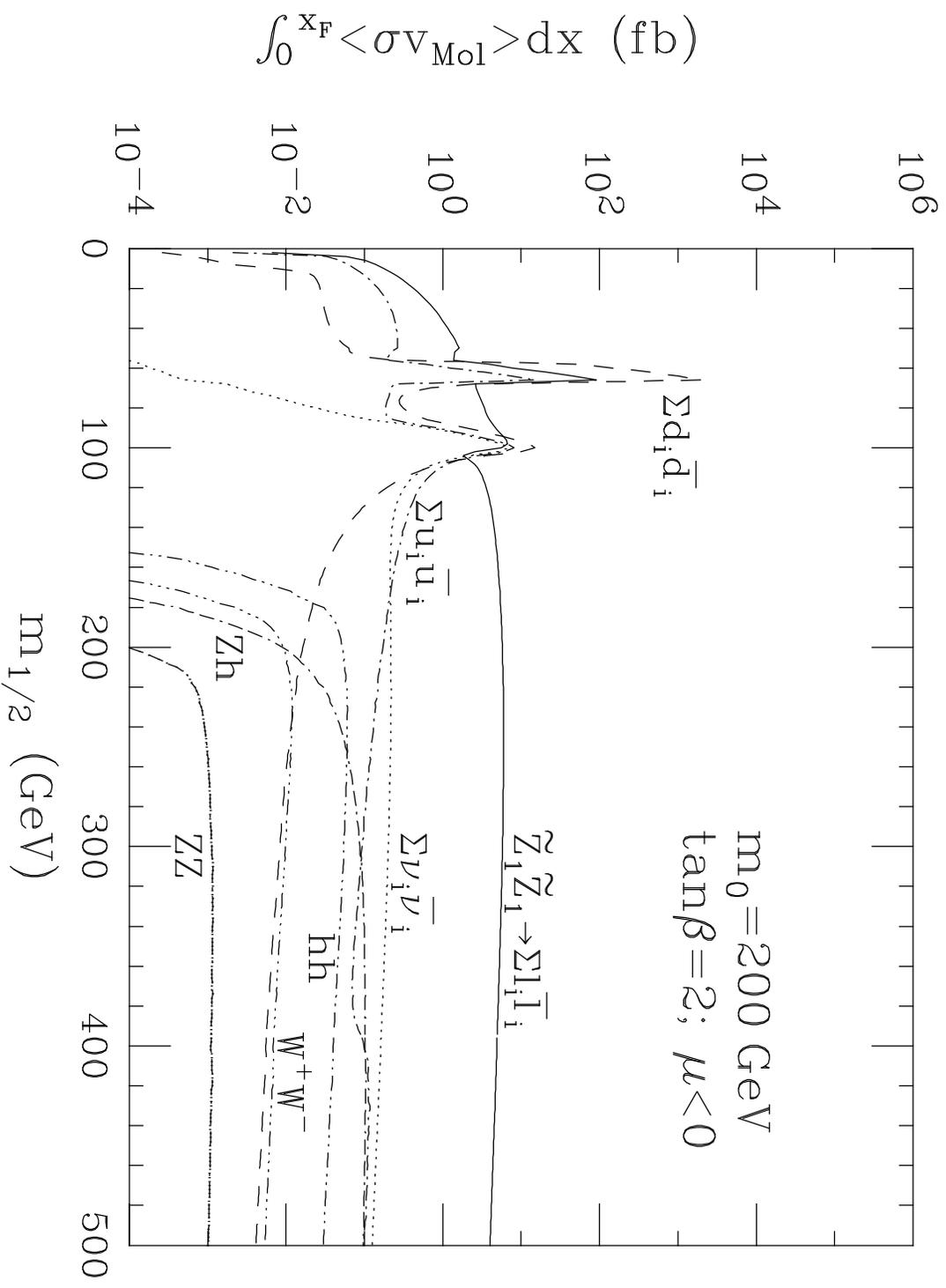
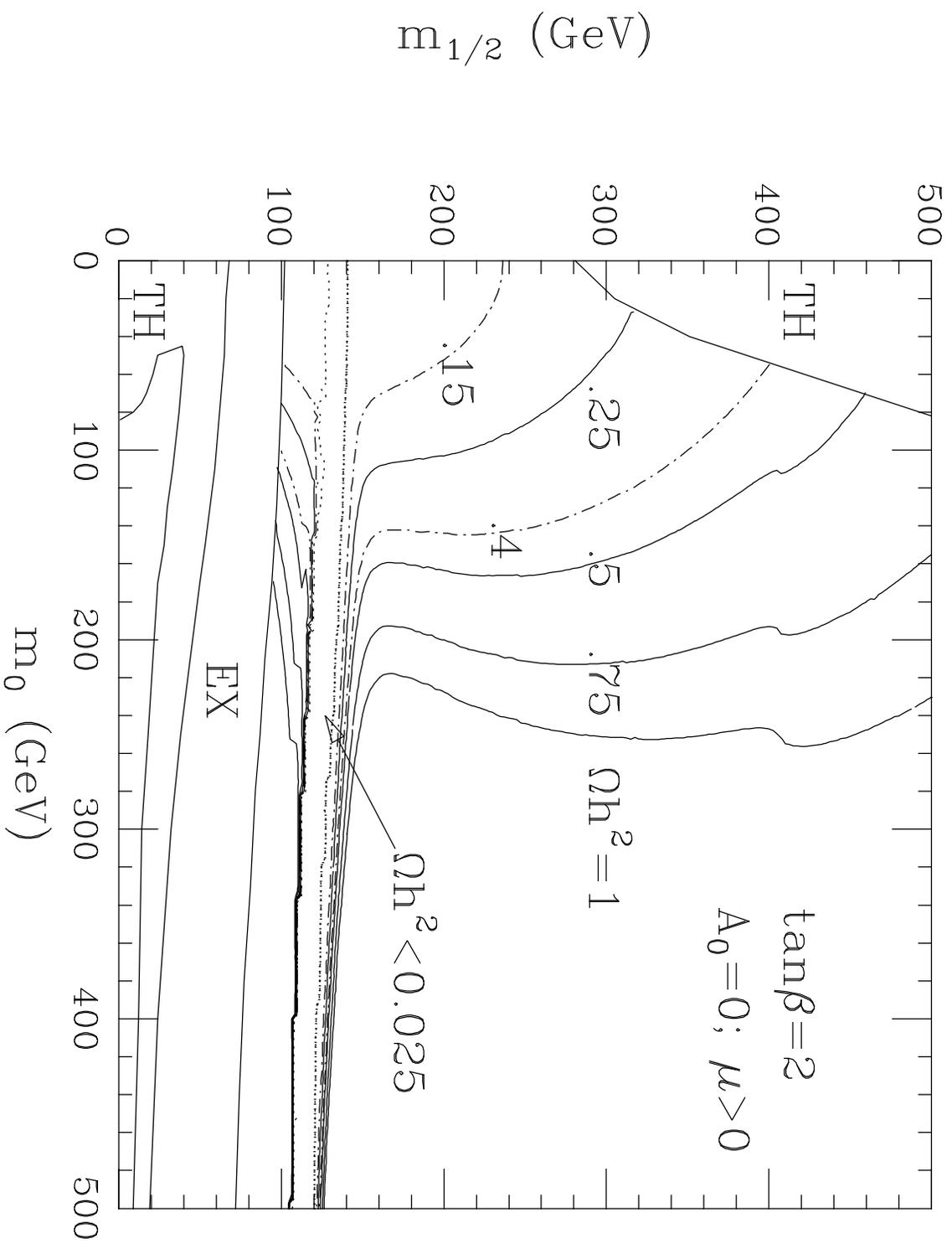
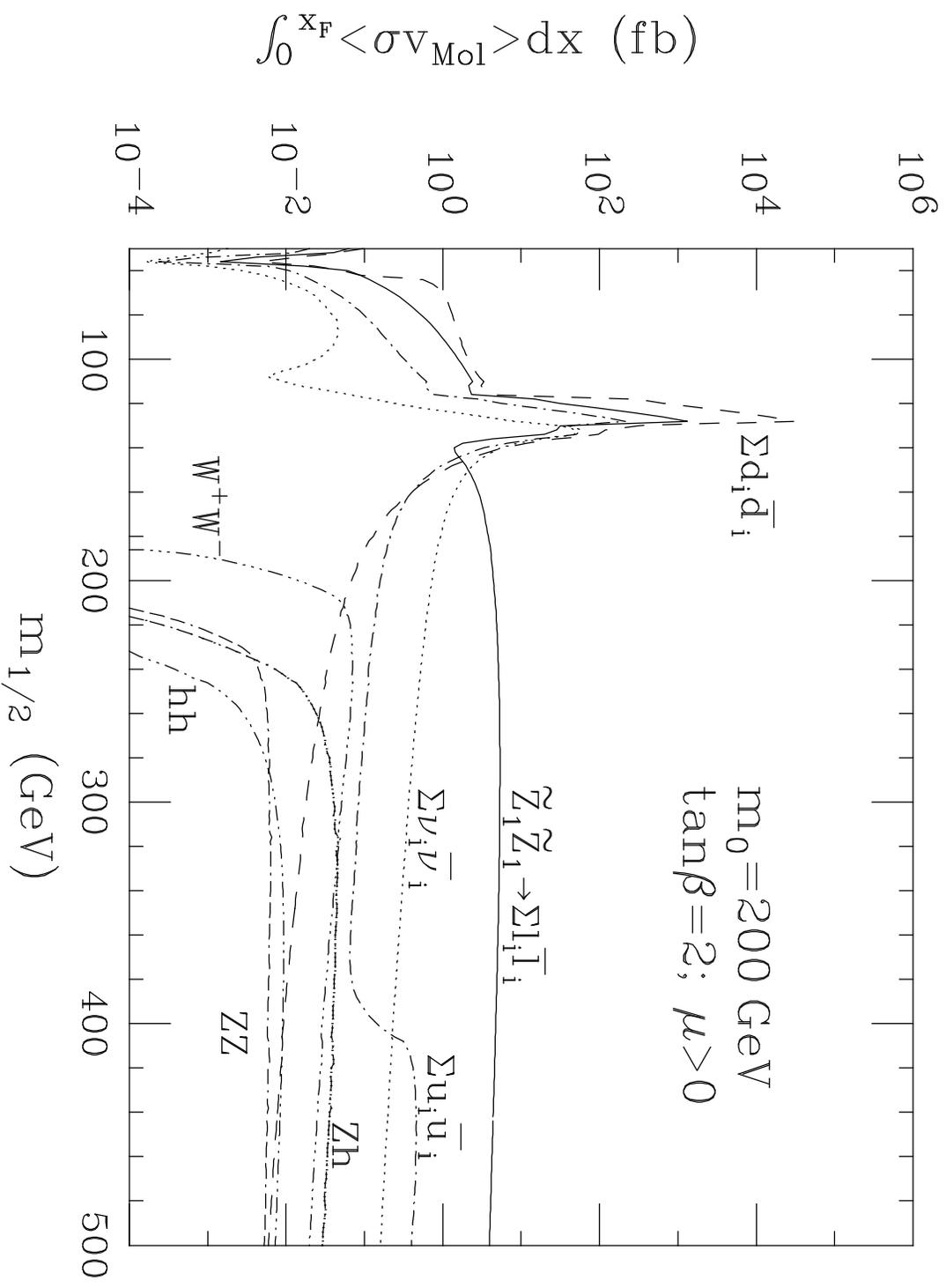
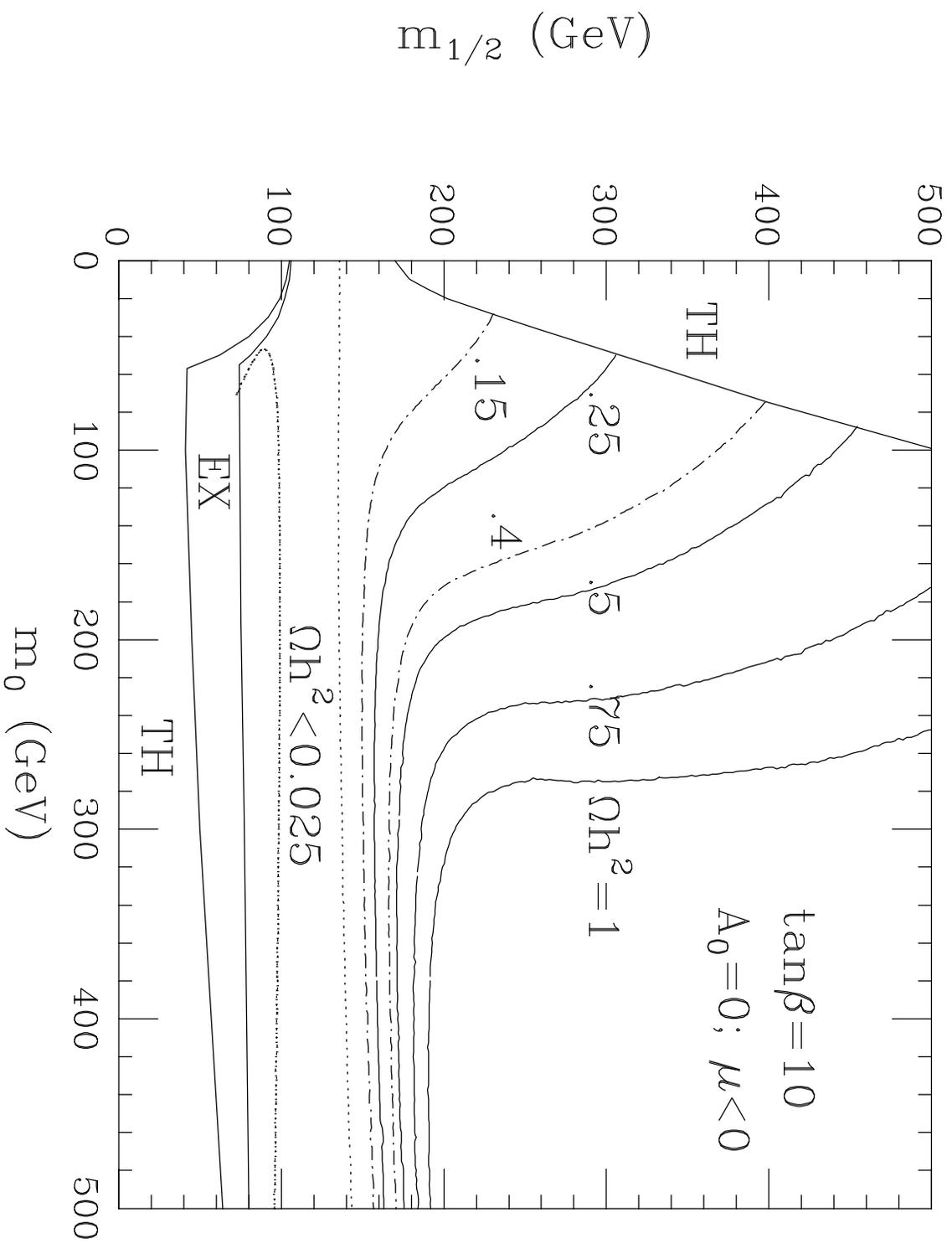
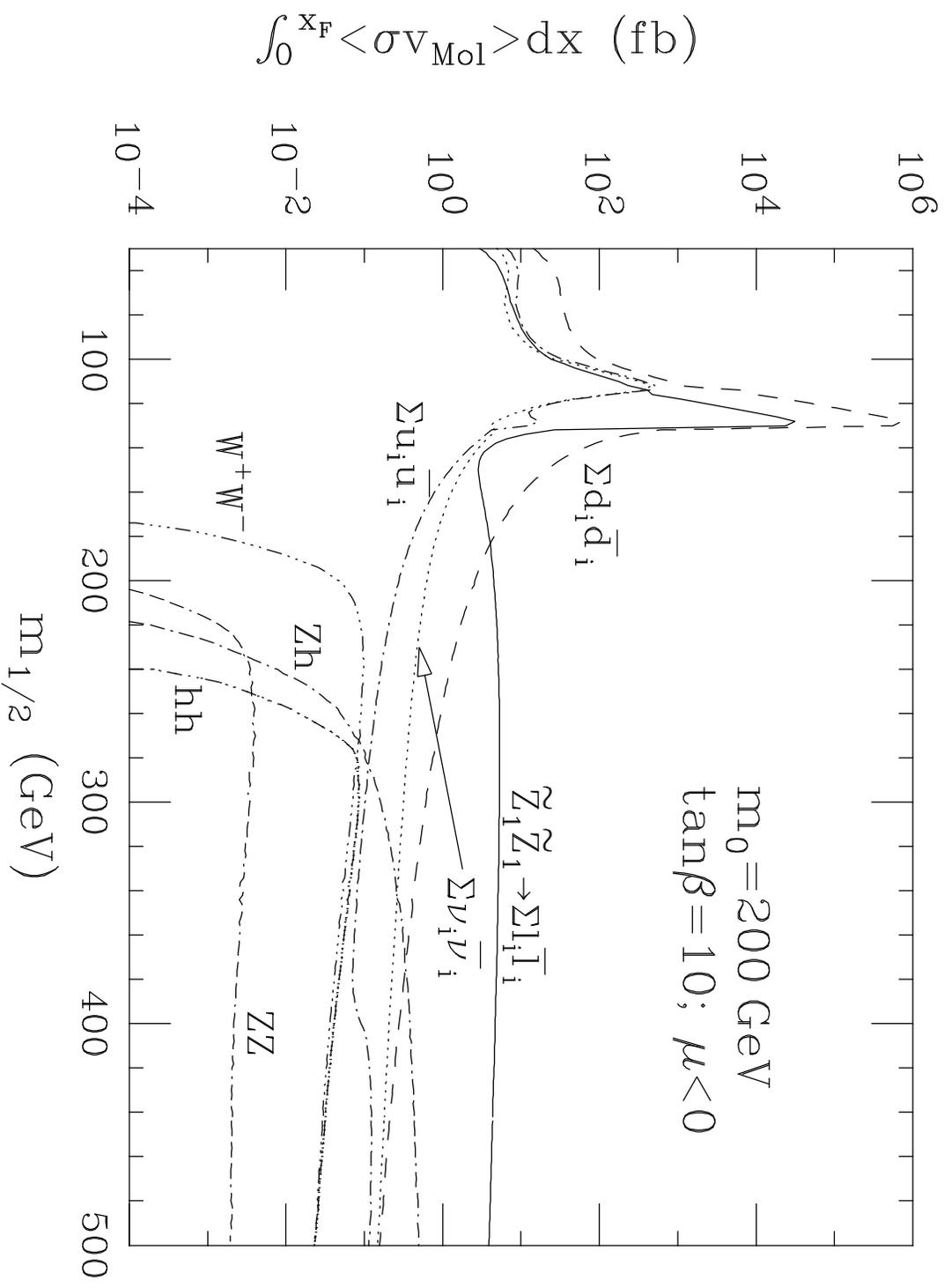
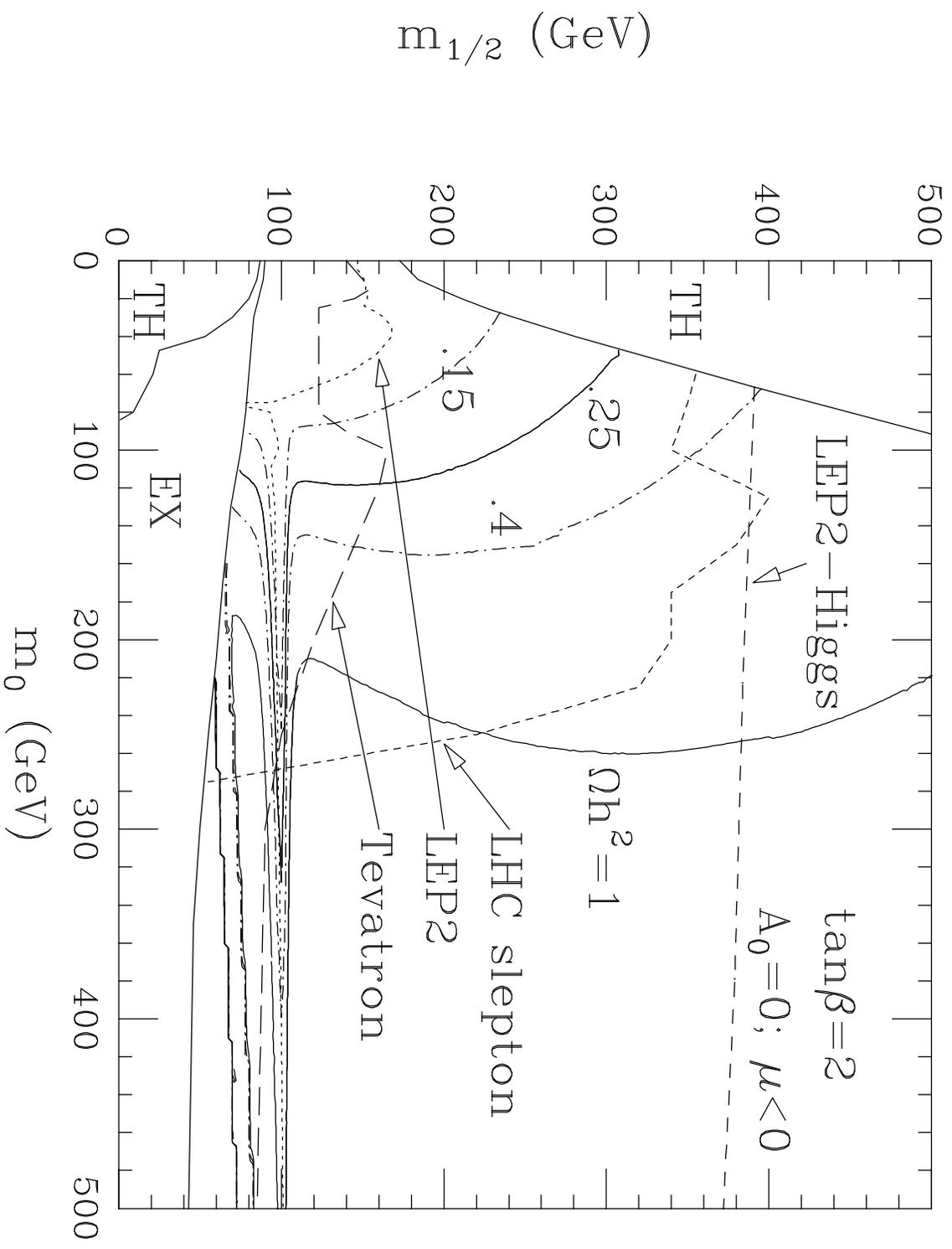
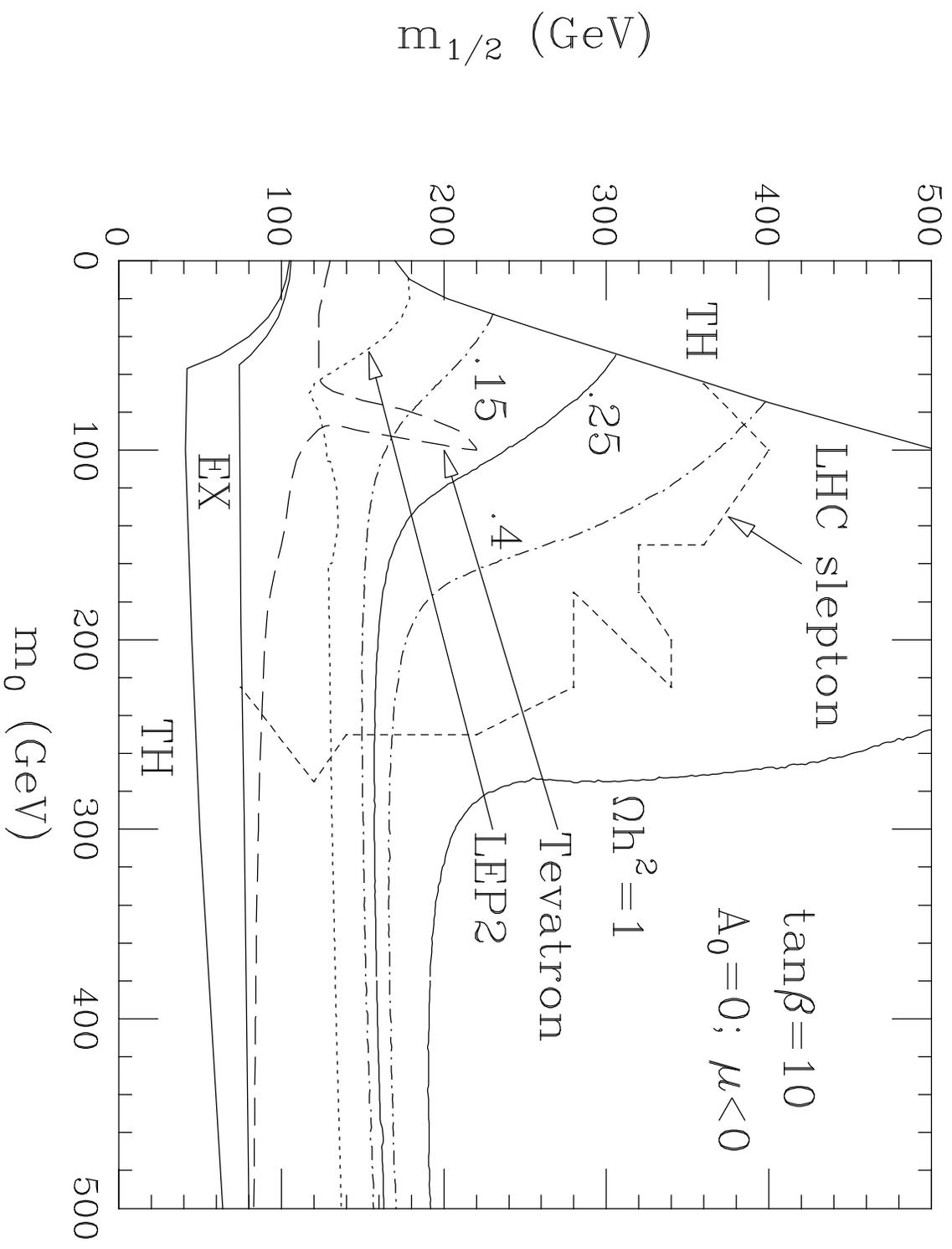
\end{document}